\documentclass{appolb}
\usepackage{epsfig}
\usepackage{amsmath,amssymb}

\newcommand{\CUDO}{{\small CUDO}}
\newcommand{\MACHO}{{\small MACHO}}
\newcommand{\MBH}{{\small MBH}}
\newcommand{\beqn}{\begin{equation}}
\newcommand{\eeqn}{\end{equation}}
\newcommand{\req}[1]{Eq.\,(\ref{#1})}

\begin{document}
\title{Compact Ultradense Objects in the Solar System%
\thanks{Presented by LL at 52 Krak\'ow School of Theoretical Physics: Astroparticle Physics in the LHC Era, Zakopane, May 19-27, 2012; \hfill DOI:10.5506/APhysPolB.43.2251 }
}
\author{Jan Rafelski$^{1}$,  Christopher Dietl$^{1}$,  and Lance Labun$^{1,2}$, 
\address{%
$^1$~Department of Physics, The University of Arizona,  Tucson, 85721 USA\\
$^2$~Leung Center for Cosmology and Particle Astrophysics, Department of Physics,\\ National Taiwan University, Taipei, Taiwan 10617
}
}
\maketitle
\begin{abstract}
We  describe  properties and gravitational interactions of meteor-mass and greater compact ultra dense objects with nuclear density or greater (\CUDO s). We discuss possible enclosure of \CUDO s in comets, stability of these objects on impact  with the Earth and Sun and show that the hypothesis of a \CUDO\ core helps resolve issues challenging the understanding of a few selected cometary impacts.
\end{abstract}
 
\section{Dark Matter and \CUDO\ s}
Considering that most matter surrounding us is yet to be discovered and its properties are at this time not known, we ask: What if this `dark' matter is not all found in free-streaming very massive particles, but a noticeable fraction is bound in  asteroid-like bodies~\cite{Labun:2011wn}?  As we will see, these bodies are not as large as stars, given the here considered high energy scale of  dark matter candidates.  For the same reason, the number of particles that need to come together to form a gravitationally bound stable body is smaller, a fraction   as small $10^{-22}$ of the number of protons in the Sun suffices.  Further, due to the high mass-energy scale, gravity dominates over other interactions even for a `small' number of particles allowing us to explore the dark asteroid structure employing well established methods~\cite{Narain:2006kx,Dietl:2011cs}. We obtain gravitationally bound objects which are naturally extremely  dense, hence  merit the name `{\bf c}ompact {\bf u}ltra {\bf d}ense {\bf o}bjects' (\CUDO s).

The question we address is how we can determine whether or not the Universe contains   \CUDO s. \CUDO s are part of the dark matter background which is explored in numerous ways today. Considering the relatively small mass of \CUDO s,   gravitational lensing cannot tell if we are observing a cloud of particles or \CUDO s or a mix. Another way is to study  visible matter dynamics employing  numerical simulations which allow for gravitating dark matter background. These simulations utilize a grainy dark matter in order to facilitate the numerical study. The \CUDO\ masses we consider are below the masses of numerical grains used, see for example Ref.~\cite{BoylanKolchin:2009nc},  and thus for these simulations presence of \CUDO s remains entirely equivalent to the effect of a dust of elementary particle dark matter.

A third way exists to observe \CUDO s and we will discuss it here: search for \CUDO s dressed in normal matter within our solar system, and collisions of \CUDO s with   planetary bodies. This in principle method  has been topic of interest in context of possible passage of a  micro-black-hole (\MBH) through the Earth~\cite{Luo:2012pp,Khriplovich:2007ci}. Such puncture collisions are also possible for the non-singular \CUDO s on account of their smallness, high energy density, and sufficient surface tidal forces~\cite{Labun:2011wn}. However, to survive  the collision with a much more massive target, a \CUDO\ must have a minimum mass in order to remain self-bound in the presence of the target attractor. 

The main difference between impact by a \CUDO\ and by a \MBH\ is that a \CUDO\  below this effective lower mass limit will dissolve and disappear in a free-steaming cloud of `dark' matter particles thus not leading to the searched for acoustic path through the Earth, and  offering another possible explanation for `evaporated' meteorite impact, which leave no significant impactor material with the surface deformation and large material stress. We address these questions in Section~\ref{CometCUD}. In the next Section~\ref{CUDOStructure} we consider solutions of Tolman-Oppenheimer-Volkoff (TOV) equations in order to characterize better \CUDO\ properties. 
\section{\CUDO\ mass and Radius}\label{CUDOStructure}
\begin{figure}
\centerline{\includegraphics[width=0.9\textwidth]{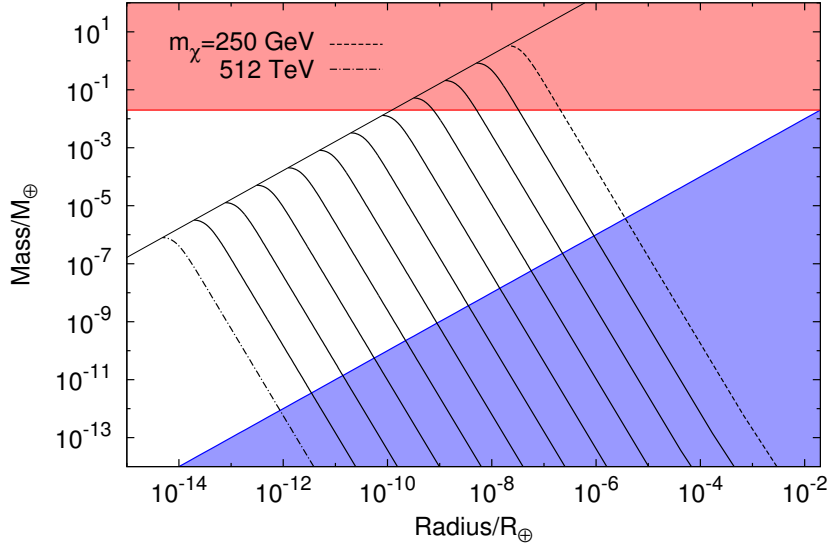}}
\caption{ Mass-radius relation (in units of Earth mass and radius)  obtained from the TOV equations  with degeneracy $g=2$ for different fermion masses: from top to bottom  $m=0.25$ up to $512$ TeV, with each successive line representing an increase in $m$ by a factor 2.  The top domain is excluded  by micro lensing surveys~\cite{Tisserand:2006zx,Carr:2009jm,Massey:2010hh}.  The bottom domain  shows where gravity of the Earth unbinds \CUDO s constituents, see \req{transfer}.
\label{MassParticle} }
\end{figure}

Paralleling the gravitationally self-bound objects composed of visible matter, two types of compact dark matter objects have been studied: those supported by Fermi-degeneracy pressure, like neutron stars, and those bounded in size by a `confining' vacuum pressure, like quark stars.  The maximum mass and corresponding radius of the gravitationally bound \CUDO\ supported by Fermi-degeneracy pressure has been determined~\cite{Narain:2006kx,Dietl:2011cs} to be  
\begin{subequations}\label{fermiMR}\begin{align}
\label{fermiM}
M_{\rm max} &= \frac{0.209}{(g/2)^{1/2}} \left(\frac{\rm 1\: TeV}{m_{\chi}}\right)^{2} M_{\oplus} \\
\label{fermiR}
R &= \frac{0.809}{(g/2)^{1/2}} \left(\frac{\rm 1\: TeV}{m_{\chi}}\right)^{2} \!\!\text{cm}=8.74\:GM_{\rm max},
\end{align}\end{subequations}
where $m_{\chi}$ is the mass of the isolated dark matter particle in vacuum and $g$ its degeneracy. We note  the scaling with inverse of the square of $m_{\chi}$.  For comparison, the mass of Earth's moon is $1.2\%\:M_{\oplus}$ and a solar mass $M_{\odot}=2.0\times10^{33}\:{\rm g}=3.3\times10^5M_{\oplus}$. A striking outcome is the  realization of extraordinary smallness, with the radius being 4.37 times the   Schwartzschild radius $R_S=2GM_{\rm max}$. The mass-radius relations obtained from numerical integration of the TOV equations are plotted in Figure~\ref{MassParticle}. 

\begin{figure}
\centerline{\includegraphics[width=0.9\textwidth]{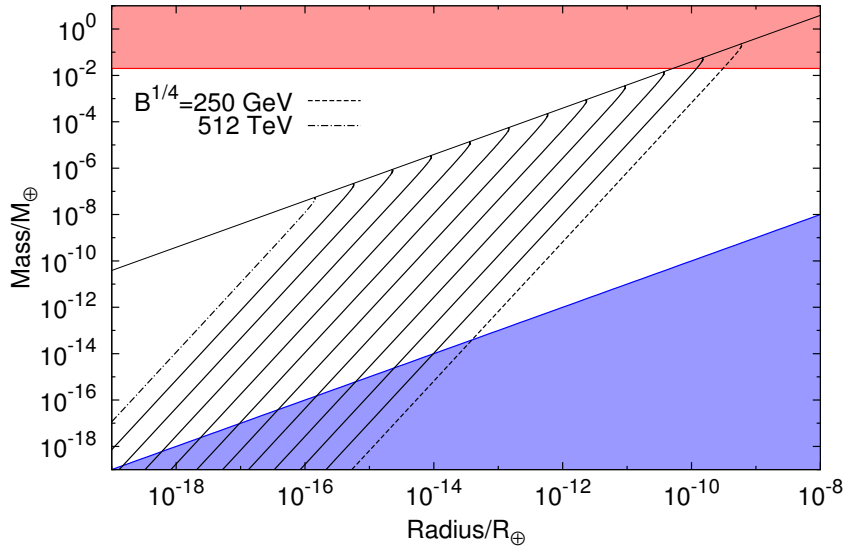}}
\caption{Same as in Figure \ref{MassParticle}, but \CUDO\ is made of  massless particles with degeneracy  $g=2$  with confining scale, from top to bottom  $B^{1/4}=0.25$ up to $512$ TeV, with each successive line representing an increase in $B^{1/4}$ by a factor 2. 
\label{MassQuark} }
\end{figure}

The quark-star analog \CUDO s have an energy scale set by the vacuum pressure (or `bag pressure') $B$.  Neglecting masses and interactions of constituent particles, the maximum mass of a structured-vacuum \CUDO~is~\cite{Lattimer:2006es}
\begin{subequations}\label{vacstarMR}\begin{align}
\label{vacstarM} 
 M_{\rm max} &= \frac{0.014}{(g/2)^{1/2}}\left(\frac{\rm 1\: TeV}{B^{1/4}}\right)^{\!2} M_{\oplus} \\
\label{vacstarR} 
 R &= \frac{0.023}{(g/2)^{1/2}}\left(\frac{\rm 1\: TeV}{B^{1/4}}\right)^{\!2}\:{\rm cm}
    = 3.69\:GM_{\rm max},
\end{align}\end{subequations}
inversely proportional to the bag pressure.  Mass-radius relations obtained from numerical integration of the TOV equations are plotted in Figure~\ref{MassQuark}.

The rising solid line in each of the figures  \ref{MassParticle} and \ref{MassQuark} corresponds to the upper mass limit defined by gravitational collapse instability, Eqs.\:\eqref{fermiMR} and \eqref{vacstarMR}.   Moving away from the instability line along a curve of particular $m_{\chi}$ or $B$ corresponds to decreasing central energy density of the \CUDO.  The shape of the curves is independent of $m_{\chi}$ or $B$.  The upper shaded domains at $M>2\times 10^{-2}M_{\oplus}$ delimits the gravitational microlensing exclusion on MACHOs: surveys show that less than 20\% by mass of dark matter in the Milky Way's halo can be found in objects with mass above this line~\cite{Tisserand:2006zx,Carr:2009jm,Massey:2010hh}.  

For a gravitationally self-bound {\CUDO} of sufficiently low mass, the potential energy at surface is small enough to allow the target-induced  polarization force to attract particles from the {\CUDO}. That mechanism in its origin is similar to the accretion of matter from one star to another in a two star system. The qualitative condition for transfer of matter across connecting path is that the presence of the target body opens a potential valley from the binding potential of the {\CUDO} at its surface $R_c$ towards the potential of the planetary target  body  at their separation. Note that both the planet and the {\CUDO} are in orbit around the Sun, so we can assume local balance of solar related dynamics and ignore the dominant but slowly varying potential of the Sun. Considering  the {\CUDO}-rocky body encounter as if occurring in free space, the {\CUDO} will impact the surface if the transfer of material from {\CUDO} to target begins only after surface penetration.  Therefore if the \CUDO\ mass satisfies
\beqn\label{transfer}
M_c>M_t\frac{R_c}{R_t},
\eeqn
it will survive up to impact at the surface and in most cases  survive transit of the target interior considering that no major change in the gravitational potential ensues.

For the Earth mass and radius for $M_t$ and $R_t$,  condition \req{transfer}  is presented as the lower rising shaded region in Figs. \ref{MassParticle} and \ref{MassQuark}.  Between the two shaded areas  we see that there is a large domain of stable \CUDO s not excluded by microlensing.  The stability domain stretches to very low  masses  ranging in figure \ref{MassQuark} down to  $10^{-19}M_{\oplus}=6\: 10^5$ kg, on account of an extremely small `atomic' size of the bound system, $10^{-17}R_{\oplus}=0.6$ \AA .

\section{Cometary \CUDO s}\label{CometCUD}
\CUDO s are massive, yet ultra-microscopic bodies. They naturally provide a gravitational condensation point in space, which can with time seed an agglomeration of matter that in general is not solid: tidal forces from other bodies may compete with the binding potential at the surface, suggesting an effective (non-volcanic) mechanism to regenerate and possibly smooth the surface.

Such odd objects seem to exist: NASA picture of the day, November 6, 2012, at http://apod.nasa.gov/apod/ap121106.html shows  the moon Methone of Saturn as photographed by the Cassini probe. It displays  a smooth surface; the expected cratering must be sub-resolution.  On this moon there must be forces refreshing the surface on a time scale smaller than the local frequency of larger impacts. A rubble-pile   held together by a central \CUDO\ would   provide a possible explanation~\cite{Egg}. 

This illustrates our  believe that most  \CUDO s  found within the Solar System would   be `dressed' with normal matter by preceding encounters with visible matter bodies. Small \CUDO s dressed in a ice rubble will have cometary appearance but  significantly enhanced impact stability.  \CUDO\ collision with rocky matter bodies results in loss of kinetic energy due gravitational tidal interaction with the impacted body~\cite{Labun:2011wn},  and consequent capture of the \CUDO\ in  the solar system.  The \CUDO\ core will practically always penetrate the target body crust,  while the ice rubble creates an impact without much residual impactor mass other than vapor and dispersed traces of cometary material.  Thus a cometary-dressed \CUDO\  will make both a meteorite-like surface impact and a puncture, but the impact damage bears an unexpected relation to the impactor mass recovered.

An example of an impactor mass which appears much greater than `observed'  is the very recent 50,000y old Canyon Diablo  Barringer Meteorite Crater, Arizona (a.k.a. Meteor Crater or Barringer Crater)~\cite{Kring2007}. Only  a small  fraction of the required impactor mass can  be accounted for: the largest 639 kg meteorite recovered is on display in the `Meteor Crater' Museum.  Models have been proposed addressing fragmentation and following  evaporation leading to melt signatures on the ground~\cite{MeloshBarringer,Artemieva:2009-11}, and in this way, the melt signature of the impact can be accounted for~\cite{MeloshBarringer}. However,  detailed   modeling of the impact could not achieve consistent description within the realm of known impactor structure~\cite{Artemieva:2009-11}.  In sum, we find in the literature no conventional matter impactor solution  for the combination of the three impact features: 1) surface impact evidence,  2) resulting impactor   material recovered and identified, and 3)  surface melt signatures. 

Considering the \CUDO\ hypothesis for  the Meteor Crater, we highlight the  conclusions of the year 2012 in Ref.~\cite{Artemieva:2009-11}: ``Any modeled scenario produces orders of magnitude more projectile material (especially, solid fragments) around the crater than historically known observations. We suggest two plausible explanations (a) the removal of these materials by erosion or by early humans; (b) a specific impact scenario involving an impactor consisting of a molten and partially vaporized jet of material (not modeled here).'' We hasten to add that there are no pre-Columbian sites of a metal-processing civilization in any reasonable distance. Moreover, to our knowledge, nobody has contemplated the erosion of a large metal mass in the Arizona desert. To us, it seems that \CUDO\ core comet supplies the necessary impactor characteristics.

An observed `cometary' impact that showed unexpected stability is the Tunguska (1908) event. Witness accounts and surface material investigations point to a Tunguska comet. In 1974,  Beasley  and Tinsley write~\cite{Beasley:1974} ``\ldots Tun\-guska catastrophe involved a body with characteristics like a cometary nucleus\ldots'', while in 2010 we read about~\cite{Kolesnikov2010}: ``Traces of cometary material in the area of the Tunguska impact (1908)''. The comet hypothesis  had not been widely accepted~\cite{Vasilyev:1998}, as it is not understood how a comet could penetrate to near the surface of the Earth.  Moreover, debate about the presence of an impact crater continues, with the most recent (May 2012) study concluding in favor of Lake Cheko as representing a small (diameter $\sim 500$ m) impact crater~\cite{Gasperini2012,Gasperini2009}, about half the diameter of the  above described Meteor Crater.  Tunguska features are compatible with the cometary \CUDO\ event properties: an icy matter surrounding the core along with a strongly gravitating central \CUDO\ body would provide the enhanced stability necessary.  On the other hand, if the mass of the central \CUDO\ is below the stability threshold \req{transfer}, then it does not survive impact with the surface, consistent with the absence of an exiting object~\cite{Beasley:1974}. 

There is a significant trail in literature of historical impacts on Earth where the suspect is a surface-impacting comet, and in some cases there is coincidence of the event with signatures of a volcanic eruption. Consider the remarkable  AD 536 event. The titles of key references speak for themselves: ``A comet impact in AD 536?''~\cite{Rigby:2004}, ``New ice core evidence for a volcanic cause of the A.D. 536 dust veil''~\cite{Larsen08}, ``South Pole ice core record of explosive volcanic eruptions in the first and second millennium AD and evidence of a large eruption in the tropics around 535 AD''~\cite{Ferris11}. A cometary \CUDO\ above the stability threshold \req{transfer} punctures the crust and simulates on exit a volcanic eruption by entraining material to upper atmosphere.  It therefore could produce the recorded subsequent cooling of the Earth using terrestrial material and hence appearing in every regard to be violent volcanic eruption, without a large associated volcano.

Collisions of  comets having \CUDO\ core with the Sun can be directly observed  and perhaps such an event is not very rare in view of gravitational focusing. A recent  mysterious and well documented case  is the survival of comet  C/2011 W3 (Lovejoy) after passing through the solar corona~\cite{Sekanina:2012gk}: ``The observed behavior (i.e. orbit, stability) is at odds with the rubble-pile (comet) model, since the residual mass of the nucleus after perihelion, estimated $\sim 10^{12}$ g (a sphere $\sim$150-200 m across), still possessed significant cohesive strength\ldots''. This observation invites a \CUDO\ gravitational core hypothesis, though efforts are made to stretch standard dynamical models far enough to explain it~\cite{Gundlach:2012fv}: ``\ldots the survival of Comet C/2011 W3 (Lovejoy) within the Roche limit of the Sun is, thus, the result of high tensile strength of the nucleus, or the result of the reaction force caused by the strong outgassing of the icy constituents near the Sun''.

\section{Conclusions} 
Each of  these examples alone would not create a case for \CUDO s. However, together these examples show a common pattern that in our opinion fits well the properties we obtained solving the TOV equations for large dark particle mass, at the level of 10's of TeV, that is beyond LHC experimental discovery reach. The interesting  feature of these solutions   is that as the scale of energy of `dark' particles increases, the maximum gravitationally stable mass decreases~\cite{Narain:2006kx,Dietl:2011cs}. We  presented this in detail in Section~\ref{CUDOStructure}, adding for the first time consideration of an effective lower mass limit. 

Gravitational collapse instability provides an upper limit on mass of \CUDO s.  For dark particle masses above a few TeV, the result of gravitational collapse would be  \MBH s  with masses below the current sensitivity limit of microlensing surveys~\cite{Dietl:2011cs}: the more general class of objects known as massive compact halo objects (\MACHO s) are ruled out for $M\gtrsim M_{\oplus}=5.97\:10^{24}\:{\rm kg}$ (i.e. larger than a fraction of the Earth's mass)~\cite{Carr:2009jm,Massey:2010hh}. Note that \MACHO s encompass  any object of sufficient mass to cause microlensing, and therefore \CUDO s  are a new member of the \MACHO\ family. Note further that rather  `conventional' \CUDO\ objects can be made of visible matter, consisting for example (strange) nuclearites, fragments of neutron stars and  even \MBH. 

An important result of the above discussion is that solar system rocky bodies (bodies with solid surfaces), e.g. Earth, Moon, Mars, Mercury, moons of Jupiter (e.g. Callisto) and large asteroids, (e.g. the protoplanet Vesta)  act as time-integrating \CUDO\ `detectors'. These targets, and even  the geologically active Earth, witness the \CUDO\ flux over  billions of years and thus at least $10^8$  times longer than the modern direct observation period. Importantly, during this time integration period the solar system samples a large peripheral domain of the Milky Way, circling the galactic center a few times and passing through spiral arm regions of dense visible matter, at locations and at a time where and when \CUDO\ flux could have been considerably higher than in our current Milky Way neighborhood.  

We presented arguments to suggest  that the \CUDO\ hypothesis represents a novel possibility in context of both present understanding of dark matter and unusual features of solar system objects.  The characteristics identified here would  perhaps  not by themselves suffice to lead to a wide acceptance of the \CUDO\ hypothesis. However, ongoing exploration within the solar system may lend further support. The presence of \CUDO\ cores in solar system  asteroid and cometary bodies results in anomalous high density, a phenomenon which is at present under investigation~\cite{Egg}. This would provide further, but still indirect, evidence. Options for direct observation will arise when  gravitometer satellites appear, such as LISA-Pathfinder~\cite{LISAPF}.\\

\noindent {\it Acknowledgments}. JR thanks Mark McCaughrean of ESA for interesting discussions. This work was supported by a grant from the U.S. Department of Energy, DE-FG02-04ER41318.
 


\begin{thebibliography}{99}
 
\bibitem{Labun:2011wn} 
  J.~Rafelski, L.~Labun, and J.~Birrell,
  ``Compact Ultra Dense Matter Impactors,''
{Phys.\ Rev.\ Lett} {\bf 110}, 111102  (2013).

\bibitem{Narain:2006kx}
G. Narain, J. Schaffner-Bielich \& I. N. Mishustin,
``Compact stars made of fermionic dark matter'',
  { Phys.\ Rev.}  D {\bf 74}, 063003 (2006).


\bibitem{Dietl:2011cs} 
  C.~Dietl, L.~Labun and J.~Rafelski,
  ``Properties of Gravitationally Bound Dark Compact Ultra Dense Objects,''
  Phys.\ Lett.\ B {\bf 709}, 123 (2012).




\bibitem{BoylanKolchin:2009nc} 
  M.~Boylan-Kolchin, V.~Springel, S.~D.~M.~White, A.~Jenkins and G.~Lemson,
  ``Resolving Cosmic Structure Formation with the Millennium-II Simulation,''
  Mon.\ Not.\ Roy.\ Astron.\ Soc.\  {\bf 398}, 1150 (2009).


\bibitem{Luo:2012pp} 
  Y.~Luo, S.~Hanasoge, J.~Tromp \& F.~Pretorius,
``Detectable seismic consequences of the interaction of a primordial black hole with Earth,''
  Astrophys.\ J.\  {\bf 751}, 16 (2012).


\bibitem{Khriplovich:2007ci}
I. B. Khriplovich, A. A. Pomeransky, N. Produit \& G.Y. Ruban, 
``Can one detect passage of small black hole through the Earth?'',
{Phys.\ Rev.\  D} {\bf 77}, 064017 (2008).



\bibitem{Tisserand:2006zx}
P. Tisserand et al. [EROS-2 Collaboration], 
``Limits on the Macho Content of the Galactic Halo from the EROS-2 Survey of the Magellanic Clouds,” Astron. Astrophys.  {\bf 469}, 387  (2007). 

\bibitem{Carr:2009jm}
B.~J. Carr, K. Kohri, Y. Sendouda, \& J.'i. Yokoyama,
``New cosmological constraints on primordial black holes'',
{ Phys.\ Rev.} {\bf D81}, 104019.1-33 (2010).

\bibitem{Massey:2010hh}
R. Massey, T. Kitching \& J. Richard,
``The dark matter of gravitational lensing''
{Rept.\ Prog.\ Phys.}\  {\bf 73}, 086901 (2010).


\bibitem{Lattimer:2006es}
J. M. Lattimer, M. Prakash, 
``Equation of state, neutron stars and exotic phases,” 
Nucl. Phys. A {\bf 777}, 479 (2006).




\bibitem{Egg}
We thank Mark McCaughrean of ESA for pointing out the remarkable structure of Methone and  the related interesting discussions also addressing study of average mass density of stellar objects and the LISA Pathfinder mission opportunities.



\bibitem{Kring2007}
D. A. Kring,
``Guidebook to the Geology of Barringer Meteorite Crater, Arizona (a.k.a. Meteor Crater)''
Lunar and Planetary Institute, LPI Contribution No. 1355 (2007).

\bibitem{MeloshBarringer}
H. J. Melosh and  G. S. Collins
``Meteor Crater formed by low-velocity impact''
Nature {\bf 434}, 157 (2005).


\bibitem{Artemieva:2009-11} 
Natalia  Artemieva, Elisabetta Pierazzo,
``The Canyon Diablo impact event: 1. Projectile motion through the atmosphere''  and
``The Canyon Diablo impact event: 2. Projectile fate and target melting upon impact''
Meteoritics \& Planetary Science {\bf  44}, 25 (2009) and {\bf 46}, 805 (2011).
 


\bibitem{Gasperini2012}
L. Gasperini,  L. Cocchi, C. Stanghellini, G. Stanghellini, F. Del Bianco, M. Serrazanetti, and C. Carmisciano, 
``Magnetic and seismic reflection study of Lake Cheko, a possible impact crater for the 1908 Tunguska Event'', 
Geochem. Geophys. Geosyst., 13, Q05008, doi:10.1029/2012GC004054 (2012).

\bibitem{Gasperini2009}
L. Gasperini, et. al. 
``Sediments  from  Lake  Cheko (Siberia),  a  possible impact  crater  for the 1908 Tunguska Event''
Terra  Nova,  {\bf 21},  489 (2009).

\bibitem{Beasley:1974}
W.H. Beasley, and B.A. Tinsley,
``Tungus event was not caused by a black hole'',
Nature, {\bf 250},   555-556 (1974)

\bibitem{Kolesnikov2010}
E.M. Kolesnikov, N.V. Kolesnikova
``Traces of cometary material in the area of the Tunguska impact (1908)''
Solar System Research, {\bf 44},  110 (2010);
in Russian: Astronomicheskii Vestnik, {\bf 44},  123 (2010).



\bibitem{Vasilyev:1998}
 N.V. Vasilyev,
``The Tunguska Meteorite problem today''
Planetary and Space Science {\bf 46}, 129 (1998).




\bibitem{Rigby:2004}
E. Rigby,  M. Symonds,  D. Ward-Thompson, 
``A comet impact in AD 536?''
Astron. Geophys. {45}, 1.23-1.26  (2004).

\bibitem{Larsen08} 
L.B. Larsen,   et al.,
``New ice core evidence for a volcanic cause of the A.D. 536 dust veil''
Geophys. Res. Lett. {35}, L04708  (2008).

\bibitem{Ferris11} 
D.G. Ferris,   et al., 
``South Pole ice core record of explosive volcanic eruptions in the first and second millennia AD and evidence of a large eruption in the tropics around 535 AD''
J. Geophys. Res.-Atmosph. {116}, D17308  (2011).




\bibitem{Sekanina:2012gk}
  Z.~Sekanina and P.~W.~Chodas,
  ``Comet C/2011 W3 (Lovejoy): Orbit Determination, Outbursts, Disintegration of Nucleus, Dust-Tail Morphology, and Relationship to New Cluster of Bright Sungrazers,''
  Astrophys.\ J.\  {\bf 757} (2012) 127
  [arXiv:1205.5839 [astro-ph.EP]].

\bibitem{Gundlach:2012fv} 
  B.~Gundlach, J.~Blum, Y.~V.~Skorov and H.~U.~Keller,
  ``A note on the survival of the sungrazing comet C/2011 W3 (Lovejoy) within the Roche limit,''
  arXiv:1203.1808 [astro-ph.EP].

\bibitem{LISAPF}
   F Antonucci, et. al.,
``The LISA Pathfinder mission,''
   Class. Quantum Grav. {\bf 29} 124014 (2012).



\end{thebibliography}
\end{document}